\documentclass[fleqn,usenatbib]{mnras}

\usepackage{newtxtext,newtxmath}
\usepackage[T1]{fontenc}

\DeclareRobustCommand{\VAN}[3]{#2}
\let\VANthebibliography\thebibliography
\def\thebibliography{\DeclareRobustCommand{\VAN}[3]{##3}\VANthebibliography}

\usepackage{graphicx}	% Including Fig.files
\usepackage{amsmath}	% Advanced maths commands
\usepackage{lipsum}  
\usepackage{float}

\newcommand{\mfive}{\ensuremath{{M}_{500}}} % m500
\newcommand{\rfive}{\ensuremath{{R}_{500}}} % r500
\newcommand{\msun}{\ensuremath{\text{M}_{\odot}}} % msun
\newcommand{\fgas}{\ensuremath{f_{\text{gas}}}} % fgas
\newcommand{\mbh}{\ensuremath{{M}_{\text{BH}}}} 
\newcommand{\mratio}{\ensuremath{{M}_{\text{BH}} / {M}_{\text{BH, median}}}}
\newcommand{\rvir}{\ensuremath{{R}_{\text{vir}}}}{}
\newcommand{\freac}{\ensuremath{{f}_{\text{re-accreted}}}}
\title[Co-evolution of Gas and BHs in FLAMINGO]{FLAMINGO: Tracing the co-evolution of hot gas and black holes in galaxy groups and clusters}

\author[E. E. Costello et al.]{
Emily E. Costello,$^{1}$\thanks{E-mail: E.E.Costello@2019.LJMU.ac.uk }
Ian G. McCarthy,$^{1}$\thanks{E-mail: I.G.McCarthy@LJMU.ac.uk }
Jaime Salcido,$^{1}$
John C. Helly,$^{2}$
Robert J. McGibbon,$^{3}$
\newauthor
Matthieu Schaller,$^{3, 4}$
Joop Schaye$^{3}$
\\
$^{1}$Astrophysics Research Institute, Liverpool John Moores University, 146 Brownlow Hill, Liverpool L3 5RF, UK\\
$^{2}$Institute for Computational Cosmology, Department of Physics, University of Durham, South Road, Durham, DH1 3LE, UK\\
$^{3}$Leiden Observatory, Leiden University, PO Box 9513, 2300 RA Leiden, the Netherlands\\
$^{4}$Lorentz Institute for Theoretical Physics, Leiden University, PO Box 9506, 2300 RA Leiden, the Netherlands
}
\date{Accepted XXX. Received YYY; in original form ZZZ}

\pubyear{\the\year{}}

\begin{document}
\label{firstpage}
\pagerange{\pageref{firstpage}--\pageref{lastpage}}
\maketitle

\begin{abstract}
The gas mass fraction of galaxy groups and clusters is a key physical quantity for constraining the impact of feedback processes on large-scale structure.  While several modern cosmological simulations use the gas fraction--halo mass relation to calibrate their feedback implementations, we note that this relation exhibits substantial intrinsic scatter whose origin has not been fully elucidated. Using the large-volume FLAMINGO hydrodynamical simulations, we examine the role of both central and satellite supermassive black holes (BHs) in shaping this scatter, probing higher halo masses than previously possible. For haloes with $\mfive < 10^{13} \msun$, we find that central BH mass correlates strongly and negatively with gas fraction, such that higher BH masses give rise to lower gas fractions at fixed halo mass, consistent with previous studies.  Interestingly, however, for $10^{13} \msun < \mfive < 10^{14.5} \msun$ the correlation reverses and becomes positive, with overmassive BHs residing in haloes with above-average gas fractions. By tracing progenitor BHs and haloes through cosmic time, we show that this behaviour is driven by the expulsion and subsequent re-accretion of halo gas, regulated by the timing of BH growth and feedback. Specifically, haloes that collapse earlier form BHs earlier, leading to earlier gas expulsion and re-accretion and a high gas fraction compared to haloes of the same present-day mass that formed later. Our results demonstrate that present-day scatter in the gas fraction–halo mass relation is strongly shaped by the early growth history of BHs and their haloes, a prediction that can be tested with future observational measurements.
\end{abstract}


%%%%%%%%%%%%%%%%%%%%%%%%%%%%%%%%%%%%%%%%%%%%%%%%%%

%%%%%%%%%%%%%%%%% BODY OF PAPER %%%%%%%%%%%%%%%%%%

\section{Introduction}
\label{sec: intro}

The gas fractions of galaxy groups and clusters provide a key observational link between the baryonic and dark matter components of the large-scale structure (LSS). In sufficiently massive clusters where the gas fractions are representative of the cosmic mean, the gas fractions can serve as a cosmological probe, constraining parameters such as the total matter density, $\Omega_\text{m}$ (e.g., \citealt{white_baryon_1993, evrard_intracluster_1997, ettori_rosat_1999, mathiesen_effects_1999, allen_cosmological_2002, ettori_are_2003, allen_constraints_2004, ettori_cluster_2009}).  In lower mass galaxy groups which are typically `missing' baryons relative to the cosmic mean, the gas fractions play a central role helping to constrain various astrophysical processes.  In particular, so-called ``baryonic feedback'' has been found to significantly modify the baryon content of haloes (e.g., \citealt{van_daalen_effects_2011, van_daalen_exploring_2020,  semboloni_quantifying_2011, mccarthy_bahamas_2017, terrazas_relationship_2020, davies_quenching_2020, schaye_flamingo_2023, vanloon_contribution_2024}). AGN feedback in the form of radiation, winds, and jets in particular can strongly affect halo gas, by heating, redistributing, or even ejecting it from the system altogether. These processes reshape both the overall gas fraction and its radial distribution, while also driving the quenching of star formation as well as regulating gas cooling and galaxy growth (for reviews see \citealt{mcnamara_heating_2007, fabian_observational_2012, gitti_evidence_2012, gaspari_linking_2020, eckert_feedback_2021}).  

The effects of AGN feedback are particularly pronounced in galaxy groups, as the energy output of AGN is comparable to (or can exceed) the binding energy of the gas (e.g., \citealt{brighenti_heated_2002, mccarthy_case_2010, gaspari_mechanical_2012}), which is why ejection leading to systematically lower gas fractions in groups is possible (see \citealt{lovisari_physical_2021} for a recent review). The mass dependence of gas fractions therefore encodes key information about the efficiency of feedback processes and their interplay with dark matter haloes.

%Note that the gas fractions of groups and clusters are typically measured via X-ray observations (e.g., \citealt{sun_chandra_2009, lovisari_scaling_2015, akino_hsc-xxl_2022, popesso_hot_2024}) and (indirectly) Sunyaev–Zel'dovich (SZ) measurements (e.g., \citealt{battaglia_future_2017, schaan_evidence_2016}, \citeyear{schaan_atacama_2021}; \citealt{mallaby-kay_kinematic_2023, hadzhiyska_velocity_2024, ried_guachalla_backlighting_2025}). 

While there is consensus that AGN feedback is a key ingredient required to produce realistic groups and clusters in cosmological simulations, it is nevertheless the case that predictions of the simulations can be very sensitive to the details of its implementation (e.g., \citealt{mccarthy_case_2010, le_brun_towards_2014, mccarthy_bahamas_2017, schaye_flamingo_2023, bigwood_case_2025}). The ability of simulations to reproduce the observed gas fractions across a wide range of halo masses has therefore become a key test of the realism of the feedback in the simulations. To this end, several semi-analytic models (e.g. \citealt{arico_bacco_2021, mead_hmcode-2020_2021, asgari_halo_2023, schneider_baryonification_2025}) and modern cosmological simulations [e.g. BAHAMAS \citep{mccarthy_bahamas_2017}; FLAMINGO \citep{schaye_flamingo_2023}] explicitly calibrate the feedback modelling to the observed median gas fraction–halo mass relation.

{We point out, however, that the gas fraction--halo mass relation exhibits significant intrinsic scatter in the simulations (e.g., \citealt{mccarthy_bahamas_2017, le_brun_scatter_2017, oppenheimer_simulating_2021, marini_impact_2025}), as well as in the observations (see e.g., \citealt{sun_chandra_2009, lovisari_scaling_2015, akino_hsc-xxl_2022, popesso_hot_2024}), suggesting that the gas fractions are shaped not only by halo mass/binding energy, but also by additional factors.  Note that it is challenging to derive robust estimates of the scatter observationally, owing to the important role that selection effects play and also uncertainties in halo mass estimation.  The study of \citet{akino_hsc-xxl_2022} may represent the best current attempt to fold in these effects for a relatively large, well-defined sample (XXL) with weak lensing-based halo mass estimates.  These authors find an intrinsic scatter of approximately 0.17 dex ($\approx50\%$) in gas mass at fixed (true) halo mass, which is comparable to the predictions of recent cosmological simulations that include the effects of AGN feedback (e.g., \citealt{le_brun_scatter_2017}).} 

In terms of what is physically driving the scatter at fixed halo mass, one possible factor, explored by \citet{davies_gas_2019, davies_quenching_2020}, is the role of assembly history in giving rise to different feedback histories for haloes of the same mass. Using the EAGLE \citep{schaye_eagle_2015} and IllustrisTNG \citep{pillepich_simulating_2018} simulations, they found a strong anti-correlation between halo gas fractions and the mass of the central supermassive black hole (BH) for haloes of mass $10^{11.5} \msun \leq M_{200} \leq 10^{13} \msun$. In other words, haloes that host more massive-than-average BHs tended to have formed their BHs earlier and to have undergone stronger AGN feedback throughout their history, which heats and expels more gas beyond $R_{200}$, leading to lower gas fractions. Conversely, haloes with undermassive BHs were found to generally retain higher-than-average gas fractions.

The FLAMINGO suite of large-scale cosmological hydrodynamical simulations \citep{schaye_flamingo_2023, kugel_flamingo_2023} provides a natural extension to this line of investigation. The suite is calibrated on both the low-z hot gas mass fractions of galaxy groups and clusters and the present-day galaxy stellar mass function (GSMF), and have large volumes of $\geq$1Gpc-a-side. This allows for robust statistics in the group and cluster regime, enabling us to probe the connection between central BH mass and gas fraction at higher halo masses than previously possible.

In this work, we begin by investigating this connection across a wide halo mass range, and explore how the growth histories of central  BHs shape the present-day scatter in halo gas fractions. Specifically, we ask \textit{when} central BHs inject feedback strongly enough to alter their host halo’s baryon content, and \textit{how} this timing influences their present-day gas fractions. By tracing the redshift evolution of haloes that host overmassive and undermassive central BHs, we aim to disentangle the physical mechanisms that drive the observed diversity in gas fractions at fixed halo mass at the present-day.

In previous literature, the focus has largely been on feedback from central BHs. However, galaxy groups and clusters form hierarchically and often host numerous additional supermassive BHs. These can reside in massive satellite galaxies (e.g., \citealt{reines_dwarf_2013}), in the central galaxy as remnants of mergers (e.g., \citealt{ricarte_origins_2021}), or exist as “free-floating” BHs ejected through gravitational slingshot interactions or tidal stripping (e.g., {\citealt{voggel_impact_2019}}). At high halo masses, the combined mass of these satellite BHs can become comparable to (or even greater than) that of the central BH. While each BH individually is typically less massive, they can still accrete and drive AGN feedback, yet their cumulative impact on halo gas has received relatively little attention. 

With this in mind, we decompose the BH populations of FLAMINGO haloes into their central and satellite components, and investigate how the satellite BH population contributes to the scatter in the gas fraction–halo mass relation across time. Our analysis reveals hints of a quenching effect exerted by early-forming central BHs on their satellites, prompting us to track fixed satellite BH populations across cosmic time and to examine the redshift evolution of radial profiles of BHs surrounding the central.

The structure of this paper is as follows. In Section \ref{sec:methods}, we give a brief description of the simulations used (with an emphasis on their BH subgrid implementations), and our methods to characterise haloes and their populations of BHs. In Section \ref{sec: examining fgas in haloes}, we examine the relationship between halo gas and halo mass, and the dependence of scatter about this relation on the mass of the various BHs within the halo, at the present-day and at higher redshift. We posit a theory that we investigate further in Section \ref{sec:Black_Hole_Tracking}, where we track the properties of a subsample of BHs and haloes through cosmic time. In Section \ref{sec:proximity effect}, we explore the idea of mutual quenching effects between central and satellite BHs, before summarising our results in Section \ref{sec:summary}.

\section{Simulations and methodology}
\label{sec:methods}
\subsection{FLAMINGO simulations}
\label{subsec: FLAMINGO Simulations}
    We provide here a summary of the key characteristics of the FLAMINGO simulations and refer the reader to \citet{schaye_flamingo_2023} and \citet{kugel_flamingo_2023} for more detailed descriptions. 

    The FLAMINGO (Full-hydro Large-scale structure simulations with Allsky Mapping for the Interpretation of Next Generation Observations) suite comprises 18 cosmological hydrodynamical simulations as described in \citet{schaye_flamingo_2023} and \citet{mccarthy_flamingo_2025}, along with two decaying dark matter variants \citep{elbers_flamingo_2025} and 12 dark-matter-only simulations. The simulations were performed using the \textsc{SWIFT} smoothed particle hydrodynamics and gravity code \citep{schaller_swift_2024}, employing the \textsc{SPHENIX} SPH scheme \citep{borrow_sphenix_2022}. Initial conditions were generated with a modified version of \textsc{monofonIC} \citep{hahn_music2-monofonic_2020, michaux_accurate_2021}, and massive neutrinos were implemented using the $\delta_f$ method from \citet{elbers_optimal_2021}.

    The FLAMINGO suite spans a range of cosmologies, box sizes, resolutions, and subgrid physics models.  Important physical processes that occur on scales too small to be directly resolved in the simulations, such as radiative cooling \citep{ploeckinger_radiative_2020}, star formation \citep{schaye_relation_2008}, stellar evolution \citep{wiersma_chemical_2009}, and feedback from stars \citep{chaikin_thermal-kinetic_2023} and AGN, are implemented with subgrid models.  The feedback processes were calibrated using machine learning-based emulators to reproduce the observed GSMF and gas fractions in groups and clusters at low redshift \citep{kugel_flamingo_2023} in the case of the `fiducial' model.  While FLAMINGO also explores variations in AGN feedback efficiency and nature (with both radiative wind and kinetic jet models), we focus on the fiducial radiative wind model in the present study.  Specifically, we use the high-resolution `m8’ run which simulates a $(1~\mathrm{Gpc})^3$ volume with $2 \times (3600)^3$ particles of mass $m_\text{g} = 1.3 \times 10^8 \msun$ (baryonic) and $m_\text{CDM} = 7.06 \times 10^8 \msun$ (dark matter). This run adopts cosmological parameters from the maximum-likelihood DES Y3 ‘3×2pt + All Ext.’ $\Lambda$CDM model \citep{des_collaboration_dark_2022} {and has a maximum proper gravitational softening length of 2.85 kpc.  At $z > 2.91$, the softening length is held constant in comoving units at 11.2 ckpc.} We make use of 13 snapshots (at redshifts $z = 0, 0.1, 0.2, 0.3, 0.4, 0.5, 0.75, 1, 1.5, 2, 3, 4$, and 5) and 78 halo catalogues, as well as a full-sky lightcone constructed on-the-fly from the hydrodynamical simulation.

\subsubsection{Black hole subgrid implementation}
\label{subsubsec: BH Implementation}
        As it is particularly pertinent to the present study, we present some additional details about the BH subgrid implementation in FLAMINGO.  The implementation is grounded in the methods of \citet{booth_cosmological_2009} (see also \citealt{springel_cosmological_2005, di_matteo_direct_2008}) and is described in detail in \citet{schaye_flamingo_2023}.
        
        {Cosmological simulations lack the resolution and detailed physics to accurately model the formation of BHs (e.g., \citealt{ volonteri_origins_2021}). Instead, in FLAMINGO BH particles are injected into haloes which surpass the mass threshold of $2.757 \times 10^{11}\msun$\footnote{{This halo mass comes from the calibration of the intermediate resolution BAHAMAS and FLAMINGO simulations, corresponding to approximately 50 particles at that resolution. Thus, essentially all `resolved' haloes are seeded with a BH at intermediate resolution.  While a higher mass threshold could have been adopted, tests demonstrate that it is difficult to match the location of the knee of the GSMF with a higher threshold for seeding.  We retain the same mass threshold for seeding in the high-resolution FLAMINGO run to enable us to match the GSMF knee and to allow a direct comparison between the intermediate and high-resolution simulations.}} (equal to the mass of 391 dark matter particles) in the case that the halo does not already contain a BH. 
        The BH particle is seeded at the position of the densest gas particle in the halo and inherits the mass of that gas particle as its \textit{particle mass}. This particle mass, typically of or beyond the order $10^8 \msun$, is a numerical quantity used solely to compute the gravitational interactions of the BH. Because this mass does not represent the unresolved physical BH mass, each BH is additionally assigned a \textit{subgrid mass} of $10^5 \msun$ at seeding. The subgrid mass is the quantity used in all sub-resolution calculations, including BH growth via gas accretion and mergers, and the associated feedback processes. Throughout this work, we therefore refer exclusively to the BH subgrid mass when discussing BH properties, and omit further mention of the particle mass. }

        BHs experience significant dynamical friction, which results in their migration to the centre of the halo and the restriction of subsequent movement. FLAMINGO lacks sufficient resolution to directly model this process, so BHs are repositioned `by hand' at each time-step of the simulation. The BH particle is relocated onto the position of the gas particle within 3 gravitational softening lengths (and within its SPH smoothing kernel) with the lowest gravitational potential energy, given that this is lower than the potential of the current position and regardless of the gas particle's velocity (as in \citealt{bahe_importance_2022}). For these calculations, the BH's own contribution to the potential must be excluded to prevent it becoming trapped in its own local potential well. This subtraction has not been applied in most earlier cosmological simulations, as well as most intermediate-resolution FLAMINGO runs. The subtraction was taken into account for the high- and low- resolution simulations as well as the intermediate-resolution kinetic AGN feedback model. Effects of this are explored in \cite{schaye_flamingo_2023} and \citet{bahe_importance_2022}. 
    
        The prescription for merging BHs follows \citet{bahe_importance_2022}, where two BH particles are merged if they are within 3 gravitational softening lengths of each other and if their relative velocities satisfy $\Delta v < \sqrt{Gm_\text{BH}/h}$, where $G$ is the gravitational constant, $m_\text{BH}$ is the mass of the larger BH and $h$ is their separation. When two BH particles are merged, momentum is conserved and the lower mass BH particle is removed from the simulation. 
    
        BH accretion is implemented following the methods of \citet{booth_cosmological_2009}, where BHs accrete at a modified Bondi-Hoyle rate dependent on the local gas density. 

        FLAMINGO implements two distinct modes of AGN feedback: the fiducial thermal model and the kinetic jets model, the latter of which was not employed in this study. The fiducial model follows the methodology of \citet{booth_cosmological_2009}, in which a fraction of the accreted rest-mass energy, given by $\epsilon_\text{r} \epsilon_\text{f} = 0.015$ (where $\epsilon_\text{r}$ is the radiative efficiency and $\epsilon_\text{f}$ is the AGN feedback coupling efficiency), is stored in a subgrid energy reservoir. Once the accumulated energy is sufficient to increase the temperature of a specified number of gas particles, $n_{\text{heat}}$, by a temperature $\Delta T_{\text{AGN}}$, it is injected into the nearest gas particles. {In FLAMINGO, $n_{\text{heat}} = 1$, and $\Delta T_{\text{AGN}}$ is a calibrated parameter of value $10^{8.07} \text{K}$ in the case of the high-resolution run (see \citealt{kugel_flamingo_2023}).  Under typical conditions at high redshift ($z\gtrsim2$), where most of the gas is ejected (\citealt{mccarthy_gas_2011}, see also Fig.~\ref{fig:BH_tracking_subplots} below), the black hole accretion is limited to the Eddington rate and the time interval between AGN feedback events is $\sim$Myr (see figure 3 of \citealt{booth_cosmological_2009}). At lower redshifts the accretion rate onto BHs in centres of groups and clusters is much lower and the typical time interval between outbursts is $\sim$100 Myr.}  We finally note that FLAMINGO recovers the $M^* -M_\text{BH}$ relation well (see \citealt{schaye_flamingo_2023}), and therefore is well-suited to this study. 

\subsection{Characterising haloes and black hole populations}

    \begin{figure}
    \centering
    \includegraphics[width=0.99\columnwidth]{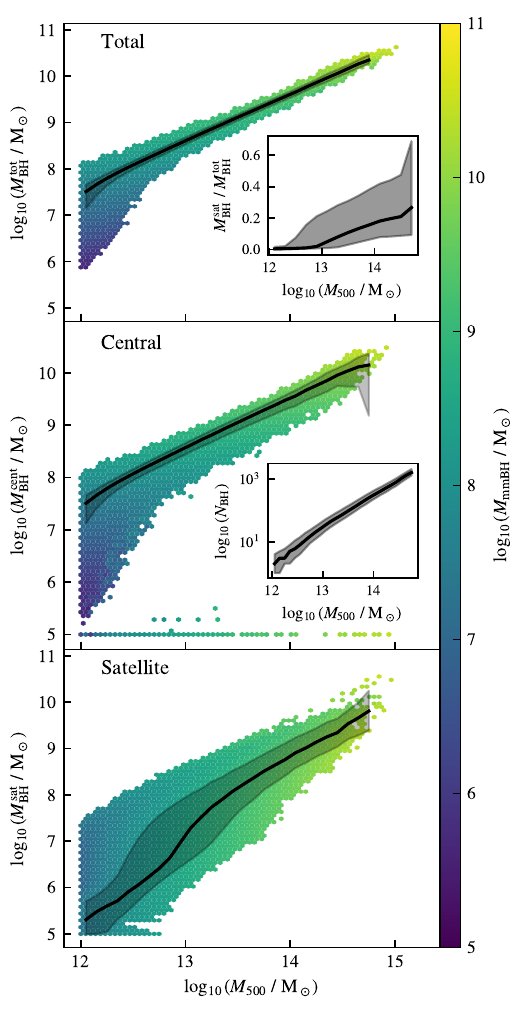}
    \caption{The present-day supermassive black hole mass, \mbh{}, as a function of halo mass, \mfive{}, for the total (top), central (centre), and satellite (bottom) BH populations in the FLAMINGO high-resolution $(1 \ \text{Gpc})^3$ volume.  Note that the satellite BH mass is defined as the sum of the masses of all satellite BHs within \rfive{}. A given hexbin contains at least three data points and is coloured by the median mass of the most massive BH (${M}_{\text{mmBH}}$) of the haloes that occupy that space in the relation. The solid black line represents the median trend and the grey shaded region encompasses the 10-90th percentile range. The inset plot in the top panel shows the median satellite BH contribution to the total BH mass budget as a function of halo mass, with the 16-84th percentile range indicated by the shaded region. {The inset plot in the middle panel shows the median number of BHs within \rfive{}, $N_{\text{BH}}$, in bins of halo mass, with the shaded region enclosing the 10-90th percentile of data.} Each population exhibits large scatter at fixed halo mass, a key driver of variations in halo gas fractions discussed later. At the most massive cluster scales, satellite BHs contribute a total mass on par with the central BH, highlighting their potential role in shaping the halo gas content.}
    \label{fig:BH_populations}
    \end{figure}
    
    Haloes are identified using the \textsc{HBT-HERONS} structure finder \citep{moreno_assessing_2025}, an updated version of the Hierarchical Bound Tracing algorithm (HBT+; \citealt{han_hbt_2018}). This structure finder begins by grouping particles with a friends-of-friends (FoF) percolation algorithm, in which a FoF group consists of dark matter particles separated by less than 0.2 times the mean interparticle distance. Gas, stellar, and BH particles in the hydrodynamical simulations are then assigned to their nearest dark matter FoF group. In addition to FoF grouping, HBT-HERONS employs a history-based approach to identify self-gravitating substructures, which are tracked from early times, with temporal information being used to disentangle complex particle distributions at subsequent times. Throughout this work, a number of halo properties across a variety of apertures are computed using the Spherical Overdensity Aperture Processor (\textsc{SOAP}; \citealt{mcgibbon_soap_2025}).

    We adopt the spherical overdensity definition of halo mass, \mfive{}, corresponding to the mass enclosed within a radius \rfive{}, where the mean interior density equals 500 times the critical density of the Universe, $\rho_\text{crit}$. Each halo is centred on its most bound particle. We select central haloes with $\mfive \geq 10^{12} \msun$, to ensure sufficient resolution. This selection yields a sample of 1,031,755 haloes, including 93,727 with $M_{500} \geq 10^{13} \msun$, and 5,378 with $M_{500} \geq 10^{14} \msun$. These large-number statistics are made possible by FLAMINGO's substantial volume. 

    We consider three types of BH population: total, central, and satellite. The mass distributions of these populations are shown in Fig.~\ref{fig:BH_populations}, where each hexbin includes at least three BHs (or BH populations). The solid black line represents the running median, computed in bins of 0.1 dex, and the shaded region indicates the 10th to 90th percentile range of the data. Each bin is colour-coded by the mass of the most massive BH (mmBH) in the corresponding halo.
    
    The total BH population (top panel) includes all BHs located within \rfive{} of their host haloes, regardless of whether they are gravitationally bound. {The number count of these BHs as a function of halo mass can be seen in the inset plot in the middle panel of Fig.~\ref{fig:BH_populations}. As expected, the number of BHs increases with halo mass, with the most massive haloes containing on the order of $10^3$ BHs.}

    {We define the central BH (middle panel) as the most massive BH within 30 kpc of the halo centre. The halo centre is defined by HBT-HERONS and SOAP as the position of the most bound particle, which may be a dark matter, stellar, gas, or BH particle. In FLAMINGO, it is not a mathematical requirement that massive BHs are repositioned onto the halo centre. Instead, BHs are repositioned onto the gas particle with the lowest potential energy within the SPH kernel of the BH particle (see Section ~\ref{subsubsec: BH Implementation}), provided that gas particle has a lower gravitational potential energy than the BH particle. As a result, BHs can in some cases reside in a local potential minimum rather than at the global halo centre (e.g., during mergers). In the majority of cases, the most massive BH migrates to the halo’s potential minimum, which typically coincides with the position of the most bound particle. However, in rare circumstances (e.g. following a major merger), these two locations can differ substantially, and the most massive BH may lie outside the 30 kpc selection radius. In such cases, our criterion misclassifies a low-mass (e.g., seed-mass) BH as the central BH. These systems appear in Fig.~\ref{fig:BH_populations} as a distinct near-seed-mass population. Each such halo does host a much more massive BH (indicated by the colour-coding), but it lies beyond the 30 kpc selection radius. Across the halo mass range studied, only 0.060\% of central BHs fall below $3 \times 10^5\msun$, and thus do not affect later results. We have examined how the position of the most bound particle changes with time for these rare cases, using the associated halo merger trees from HBT-HERONS. We find that these rare cases tend to show rapid evolution of the location of the most bound particle compared to the much more typical case of a smoothly varying position (due to the cluster bulk velocity), which would indeed suggest that there have been recent dynamical disturbances for these rare cases.} 

    We define satellite BHs as all BHs within \rfive{} of the central halo, excluding the central BH itself. This definition therefore includes BHs in satellite galaxies as well as free-floating BHs, ejected or displaced through processes such as tidal stripping or gravitational slingshot interactions. The bottom panel of Fig.~\ref{fig:BH_populations} shows the \textit{total} mass of all satellite BHs per central halo, rather than individual subhalo BHs. Compared to centrals, these populations display much larger scatter, reflecting their less self-regulated accretion and highly diverse merger histories. The inset plot in the top panel demonstrates the median contribution from satellite BHs to the total BH mass budget as a function of halo mass. For the most massive clusters, the combined satellite BH mass becomes comparable to that of the central, corresponding to $\approx50\%$ of the total BH mass budget\footnote{Note that this may be an underestimate, as the repositioning scheme used in FLAMINGO likely maximises the merger rates of BHs.}. Despite this, the cumulative feedback from satellite BHs has received little attention in the literature, motivating our investigation into their role alongside central BHs.

    As expected, the masses of both central and satellite BHs increase with halo mass, with each population exhibiting substantial intrinsic scatter (0.24 dex for central BHs and 0.56 for satellites in haloes with $10^{13.75} \msun \leq \mfive < 10^{14} \msun$). A key question is whether this scatter couples to baryonic properties of the halo, given that BH feedback can reshape the distribution and thermodynamic state of halo gas. We therefore now turn to examine the dependence of the halo gas fraction on BH mass.

\section{Gas fraction--halo mass relation: dependence on BH mass}
\label{sec: examining fgas in haloes}

\begin{figure}
    \centering
    \includegraphics[width=\columnwidth]{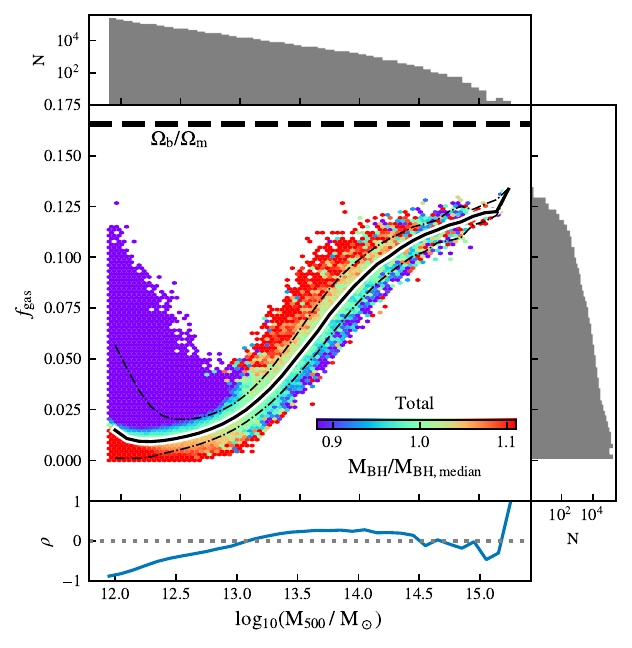}
    \caption{The present-day relation between halo gas fraction, \fgas, and halo mass, \mfive. Each hexbin is colour-coded by the median BH mass ratio, \mratio, for the total BH population. This quantifies the scatter in BH mass by taking the ratio between a halo's total BH mass and the median total BH mass at fixed halo mass. The dot-dash lines enclose the 10-90th percentile range. The colour bar limits correspond to the 16th–84th percentile range of median \mratio{} values for haloes with $\mfive \geq 10^{12.5} \msun$. The solid line shows the running median of \fgas{} in bins of 0.1 dex in \mfive, while the horizontal dashed black line represents the universal baryon fraction in FLAMINGO. Marginal histograms of \fgas{} and \mfive{} are displayed to the right of and above the main panel, respectively. The Spearman rank correlation coefficient, $\rho$, quantifying the \fgas-\mratio{} relation, is indicated below the main panel. This relation exhibits substantial intrinsic scatter, which is reduced following the onset of AGN feedback. At low halo masses, a clear negative correlation emerges: overmassive BH populations reside in haloes with below-average gas fractions, and vice versa. Above $\mfive \approx 10^{13} \msun$, however, the trend reverses, with overmassive BH populations instead associated with haloes of higher-than-average gas fractions.}
    \label{fig:tot_hex}
\end{figure}

    We begin by examining the relationship between the halo gas fraction and halo mass at the present day, and the dependence of scatter about this relation on the mass of the BHs within the halo. The gas fractions of haloes were computed as the ratio of gas mass within \rfive \space to the total halo mass, \mfive, regardless of temperature, i.e., $\fgas = M_{\text{gas}}(r < \rfive)/\mfive$. To quantify the scatter in BH mass across the halo population, for each BH we compute the ratio of its mass to the median BH mass at fixed halo mass.  These ratios are calculated in bins of 0.1 dex separately for the total BH population, as well as separately for centrals and satellites. Note that the ratio $M_{\text{BH, satellite}}/M_{\text{BH, satellite median}}$ excludes haloes with zero satellite BH mass. 

    Fig.~\ref{fig:tot_hex} shows the present-day halo gas fraction, \fgas, as a function of halo mass, \mfive. The solid line shows a running median, computed in bins of 0.1 dex, the dot-dashed lines indicate the 10-90th percentile range, and the horizontal dashed line gives the universal baryon fraction in FLAMINGO. The Spearman rank correlation coefficient, $\rho$, for the $\fgas-\mratio$ relation is shown beneath the main panel for the total BH mass. Most scatter in the gas fraction occurs at low halo masses, below $\mfive \approx 10^{12} \msun$. In this regime, haloes have relatively shallow potential wells, and gas expulsion is primarily driven by supernova feedback (see \citealt{schaye_flamingo_2023} for details of the stellar feedback prescription). AGN feedback becomes effective at higher halo masses, leading to a sharp reduction in the gas fraction and its scatter. This is also where we see the trend transition from low-mass gas-poor galaxy haloes to high-mass gas-rich galaxy group ($10^{13} \msun \leq \mfive < 10^{14} \msun$) and cluster ($\mfive \geq 10^{14} \msun$) haloes. Beyond this mass scale, the scatter remains approximately constant before declining at $\mfive \approx 10^{14.5} \msun$, where the mass content is more reflective of the cosmic mean. The trend begins to plateau below the cosmic baryon fraction in the galaxy cluster regime, the deficit accounted for by the baryons within stars. 

    The large volume of FLAMINGO enables significantly improved halo statistics compared to previous studies, particularly at the high-mass end. This is illustrated in the marginal histograms (top and right panels of Fig.~\ref{fig:tot_hex}), which show the number counts of haloes across the relation.
    
    In Fig.~\ref{fig:tot_hex}, each hexbin is colour-coded by the median \mratio{} for the total BH population. The colour bar shown within the main panel has limits corresponding to the 16-84th percentile range for all hexbins for haloes above $\mfive \geq 10^{12.5} \msun$. In other words, red (blue) hexbins indicate haloes hosting overmassive (undermassive) total BH populations relative to the median total BH mass typical for their halo mass. We see a strong negative correlation at low halo mass, $\mfive \approx 10^{12} \msun$, as reflected in the Spearman coefficient. Beyond this mass scale, the correlation strength decreases until it becomes weakly positive at $\mfive \approx 10^{13} \msun$. To further investigate this behaviour, we next decompose the total BH population into its central and satellite components.

    \begin{figure}
        \centering
        \includegraphics[width=\columnwidth]{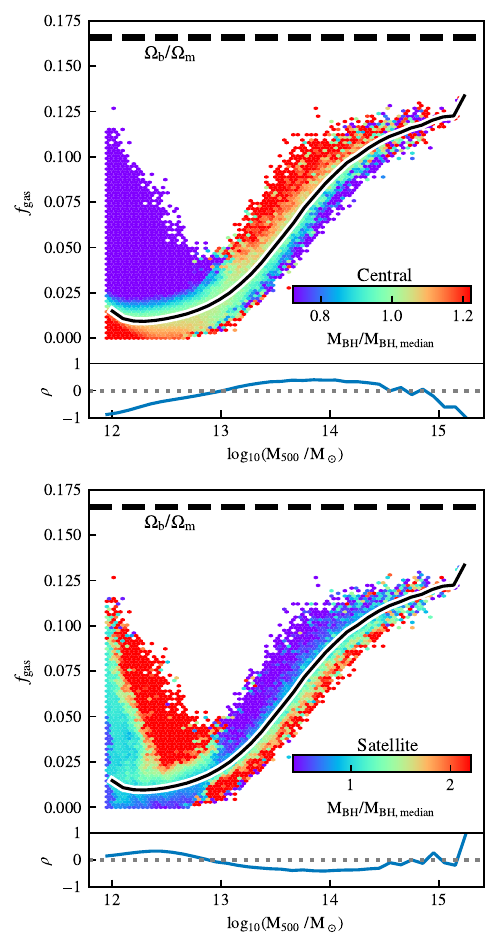}
        \caption{The present-day $f_{\text{gas}}-\text{log}_{10}(\mfive)$ relation colour-coded by the median \mratio{} for the central (top) and satellite (bottom) BH populations. The colour bar limits are set by the interquartile range of hexbin medians for haloes above $\mfive \leq 10^{12.5} M_{\odot}$ for each BH population. The black solid line indicates the running median, the horizontal dashed black line represents the universal baryon fraction in FLAMINGO. The running Spearman rank correlation coefficient is displayed below each panel. Consistent with previous work, haloes with $\mfive \lessapprox 10^{12.5} \, \msun$ that host overmassive central BHs tend to have below-average gas fractions, and vice versa. Beyond this halo mass, however, the correlation becomes positive, where overmassive central BHs reside in haloes with \textit{higher} gas fractions. The opposite trend is evident in satellite BHs, suggesting an anti-correlation between central and satellite BH mass: haloes with overmassive central BHs are likely to have an undermassive population of satellite BHs.}
        \label{fig:cent_sat_vertical_hex}
    \end{figure}

    \begin{figure*}
        \centering
        \includegraphics[width=\textwidth]{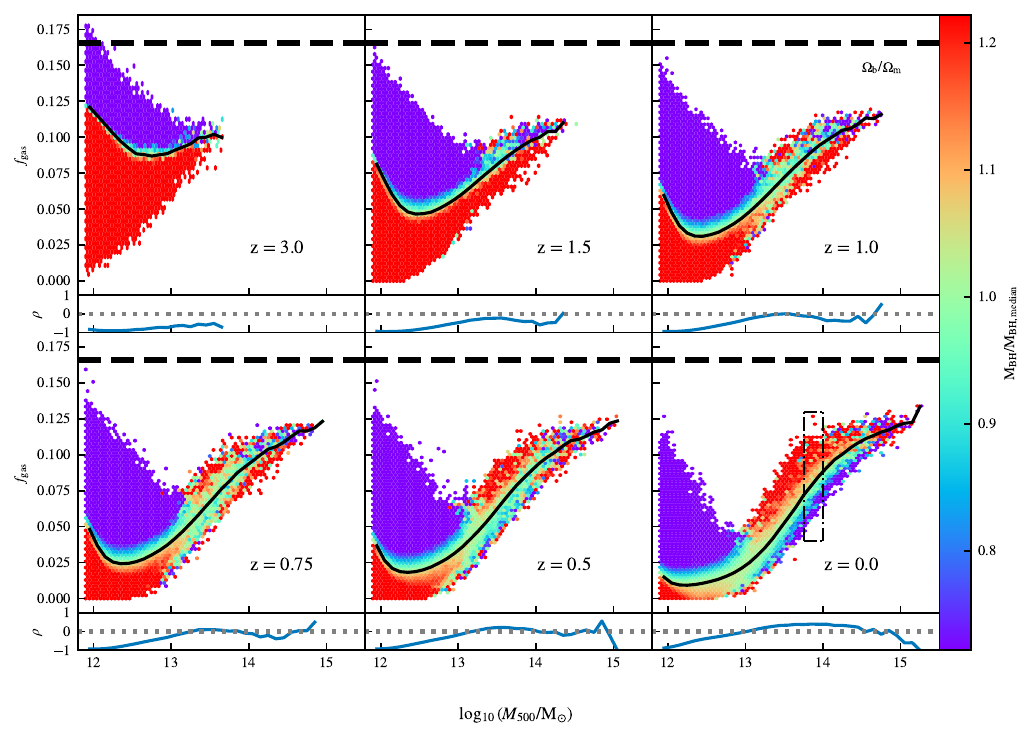}
        \caption{{Relations between halo gas fraction, \fgas, and halo mass, \mfive, across the redshift range $3 \leq z \leq 0$, colour-coded by the median \mratio{} for central BHs at that redshift. The maximum and minimum colour-bar limits are set by the interquartile range of hexbin medians for haloes with $\mfive \geq 10^{12.5} \msun$ at $z = 0$.} The solid black curves show the running median, calculated in bins of 0.1 dex. The corresponding Spearman rank correlation coefficient, calculated in bins of 0.1 dex, is shown below each main panel. The horizontal dashed line shows the universal baryon fraction in FLAMINGO, the dot-dashed enclosed region in the bottom right panel shows haloes within the mass bin $10^{13.75} \msun \leq \mfive < 10^{14}\msun$, whose BHs we track through cosmic time in Section~\ref{sec:Black_Hole_Tracking}. For $z > 1.5$, a strong negative correlation is evident, with all overmassive BHs residing in haloes with low gas fractions, independent of halo mass. For $0.25 \leq z \leq 1.5$, this trend transitions, as an increasing fraction of massive haloes with overmassive central BHs exhibit elevated gas fractions. By $z=0$, haloes above $\mfive \approx 10^{13}\msun$ predominantly host overmassive central BHs and have above-average gas fractions. In Section~\ref{sec:Black_Hole_Tracking}, we show that this reversal is driven by the ejection and subsequent re-accretion of halo gas.}
        \label{fig:multiz_horizontal}
    \end{figure*}

    Fig.~\ref{fig:cent_sat_vertical_hex} presents the $\fgas$–$\mfive$ relations at $z=0$, colour-coded by \mratio\ for both the central (top) and satellite (bottom) BH populations. As in the previous figure, the solid black line indicates the running median, the horizontal dashed line represents the cosmic baryon fraction, and the sub-panel beneath each main plot shows the running Spearman rank correlation coefficient for the \fgas-\mratio{} relation for the corresponding BH population. We begin by directing the reader's attention to the upper panel.
    
    \citet{davies_gas_2019, davies_quenching_2020}, in their studies of L* galaxies in EAGLE \citep{schaye_eagle_2015} and IllustrisTNG \citep{pillepich_simulating_2018}, demonstrated that overmassive central BHs are associated with lower-than-average gas fractions across the approximate halo mass range $10^{11.5}\msun \leq M_{200} \leq 10^{12.5} \msun$. This result is physically intuitive, as more massive central BHs are expected to have injected more feedback energy, which can expel gas more efficiently, thereby reducing the gas fraction. A similar trend is evident in FLAMINGO, as seen in the top panel of Fig. \ref{fig:cent_sat_vertical_hex}, where haloes with $\mfive \lessapprox 10^{12.5} \msun$ hosting overmassive central BHs tend to have below-average gas fractions, and vice versa.
    
    However, at higher halo masses, which are not well-sampled by smaller volume simulations like EAGLE, we observe an intriguing reversal of this trend: overmassive BHs are now associated with \textit{higher} gas fractions. The inversion is reflected in the Spearman correlation coefficient, although the quantitative strength of the correlation appears weaker than suggested by visual inspection. This apparent discrepancy may arise because the plots do not weight BHs by their number density per bin. By construction, BHs tend to cluster around the median mass ratio $(\mratio = 1)$, but our focus here is on BHs that deviate significantly from the median and here the statistics are more limited.
    
    At the highest halo masses, in the galaxy cluster regime ($10^{14.5}\msun < \mfive \lessapprox 10^{15} \msun$), the correlation weakens further. This is due to a combination of poorer statistics and the increasing gravitational binding energy of haloes, which reduces the efficiency of BH feedback in removing gas from the system.

    Turning to the bottom panel of Fig.~\ref{fig:cent_sat_vertical_hex}, which shows the satellite BH populations, we first note that low-mass haloes with low gas fractions tend to host satellite BHs with a lower-than-average combined mass. Interestingly, this correlation reverses at $M_{500} \approx 10^{13} \msun$ where high-mass haloes with high gas fractions are found to host undermassive satellite BH populations. This behaviour appears to be the inverse of the trend observed for central BHs; that is, haloes with overmassive central BHs are typically associated with undermassive satellite BH populations, and vice versa. We emphasise that this anticorrelation is not merely a consequence of the conservation of total BH mass (where ${M}_\text{BH}^\text{tot} = {M}_\text{BH}^\text{cent} + {M}_\text{BH}^\text{sat}$), as each \mratio{} was computed independently, with respect to their own population.  For example, because there is scatter in the total BH mass, it is possible for the same halo to simultaneously have both overmassive (or undermassive) central and satellite BH masses with respect to the median central and satellite BH masses at that halo mass.  Having said that, haloes where satellite BHs are able to more successfully merge with the central BH may contribute to the anticorrelation observed in the bottom panel of Fig.~\ref{fig:cent_sat_vertical_hex}.  Alternatively, it may be the result of preprocessing of satellite galaxies prior to or during their infall, or that feedback from the central BH directly influences the growth of surrounding satellite BHs.  

    Comparing this figure to Fig.~\ref{fig:tot_hex}, we find that central BHs dominate the overall trend across most of the halo mass range. However, there is some evidence that satellite BHs contribute more significantly at the highest halo masses. This is expected, as massive haloes host a larger number of satellite galaxies, and the combined mass of their BH populations can become comparable to, or greater than, that of the central BH.

\subsection{Redshift evolution}

    To investigate the surprising reversal in the \fgas-\mratio{} correlation, it is helpful to examine its redshift evolution. This may help determine whether the reversal emerges at a specific epoch in the simulation, and whether or not this coincides with key physical transitions, such as the onset of efficient AGN feedback, the establishment of hot gaseous haloes, or the re-accretion of previously expelled gas.
    
    {Fig.~\ref{fig:multiz_horizontal} illustrates the redshift evolution of the relationship between halo gas fraction and central BH mass ratio. Similar to Fig.~\ref{fig:cent_sat_vertical_hex}, Fig.~\ref{fig:multiz_horizontal} shows the halo gas fraction as a function of halo mass, colour-coded by the median central BH mass ratio, \mratio. Each panel corresponds to a different redshift within the range $0 \leq z \leq 3$. The colour scale reflects the 16th–84th percentile range of median central \mratio\ values at $z=0$, and is applied uniformly across all panels for consistent visual comparison. Again, the solid black curve denotes the median relation between gas fraction and halo mass, while the horizontal dashed black line indicates the universal baryon fraction. The Spearman rank correlation coefficient, calculated in bins of 0.1 dex, is shown below each panel. }

    As cosmic time progresses, haloes grow hierarchically and are able to reach higher masses; consequently, at high redshift, the most massive haloes (i.e., massive galaxy groups and clusters) are not yet in place and thus are absent from the sample. In the top leftmost panel ($z=3$), a strong negative correlation is evident: haloes hosting overmassive (undermassive) central BHs exhibit systematically lower (higher) gas fractions, largely independent of halo mass.
    As the simulation evolves, the onset of a reversal in this correlation can first be seen at approximately $z \approx 1.5$, where some haloes with overmassive BHs begin to retain or reacquire higher gas fractions. By $z=1$, halo growth extends the sampled mass range, and the correlation continues to weaken. At this stage, haloes with $M_{500} \sim 10^{13}$–$10^{14} \msun$ hosting overmassive BHs now appear both above and below the median gas fraction. In contrast, more massive haloes ($M_{500} \sim 10^{14}$–$10^{15} \msun$) with overmassive BHs largely retain suppressed gas fractions. By $z=0$ (bottom rightmost panel), the correlation has fully flipped. Haloes with $M_{500} \gtrsim 10^{13}~\msun$ now predominantly exhibit high gas fractions in the presence of overmassive central BHs, indicating a reversal in the nature of the coupling between BH growth and halo gas content over time.

    We posit that a possible explanation for the observed change in correlation is the recapture of previously ejected gas.  Haloes hosting overmassive central BHs at low redshift will have typically formed their BHs early in cosmic history. These BHs will have grown via a combination of accretion and mergers, eventually generating powerful feedback that expels large amounts of gas from the halo, and prevents the BH's own growth via accretion. However, while accretion is suppressed, the central BH may continue to grow through mergers as the host halo evolves. Gas that was expelled (but not far enough to escape the halo's potential well) during earlier feedback episodes can later be re-accreted onto the halo. The gas content of the halo can also increase through feedback events from nearby systems by redistributing their gas into the Lagrangian region that will eventually become the cluster. This will result in haloes that host overmassive central BHs and simultaneously exhibit higher-than-average gas fractions at late times. We explore this hypothesis in more detail in Section~\ref{sec:Black_Hole_Tracking}, where we track individual progenitor BHs and their associated halo properties across cosmic time.

\section{Co-evolution of black holes and gas}
\label{sec:Black_Hole_Tracking}  
    As outlined in the previous section, high-mass haloes hosting overmassive BHs begin to show rising gas fractions around $z\approx1.5$. To assess whether this trend is directly tied to BH mass, we track individual progenitor central BHs and connect them to their corresponding halo properties across redshift.
    
    We begin by selecting group-scale haloes at $z = 0$ within the mass range $10^{13.75} \msun \leq \mfive < 10^{14} \msun$, corresponding to the regime where overmassive central BHs are associated with high gas fractions, and vice versa. This mass bin is highlighted in a dash-dot box in Fig. \ref{fig:multiz_horizontal}. From this set, we identify haloes above and below the 90th and 10th percentiles of the gas fraction distribution, which we refer to as the `upper' and `lower' \fgas{} samples, respectively. As noted, for each selected halo we identify the central BH as the most massive BH within a 30~kpc radius of the halo centre, yielding a total of 1420 central BHs, evenly split between the upper and lower samples.

    Because only a limited number of high-resolution snapshots were stored long-term, we combine both lightcones and snapshots to track BHs and their host haloes through cosmic time. At high redshift, where snapshots are unavailable, we use halo and particle lightcone outputs, linking BH particles to their nearest halo centre with the \texttt{cKDTree} distance search algorithm from the \texttt{scipy.spatial} module, and retrieve halo properties via SOAP. At low redshift, snapshots cover the full simulation volume, and we therefore use them directly to track BHs, again using the SOAP catalogues to access halo properties.

    To further investigate the anti-correlation between the central and satellite BH masses seen in Fig. \ref{fig:cent_sat_vertical_hex}, we apply a similar procedure to track the evolution of satellite BHs. We select the same halo sample as before, again focusing on those in the upper (>90th percentile) and lower (<10th percentile) extremes of the gas fraction distribution in massive groups. From each of these haloes, we identify all BHs within $R_{500}$, excluding the central BH. This fixed set of progenitor satellite BHs, defined at $z=0$, is then traced back in time using the same procedure as before, with properties grouped by redshift and host halo ID at $z=0$.

\begin{figure}
 \includegraphics[width=\columnwidth]{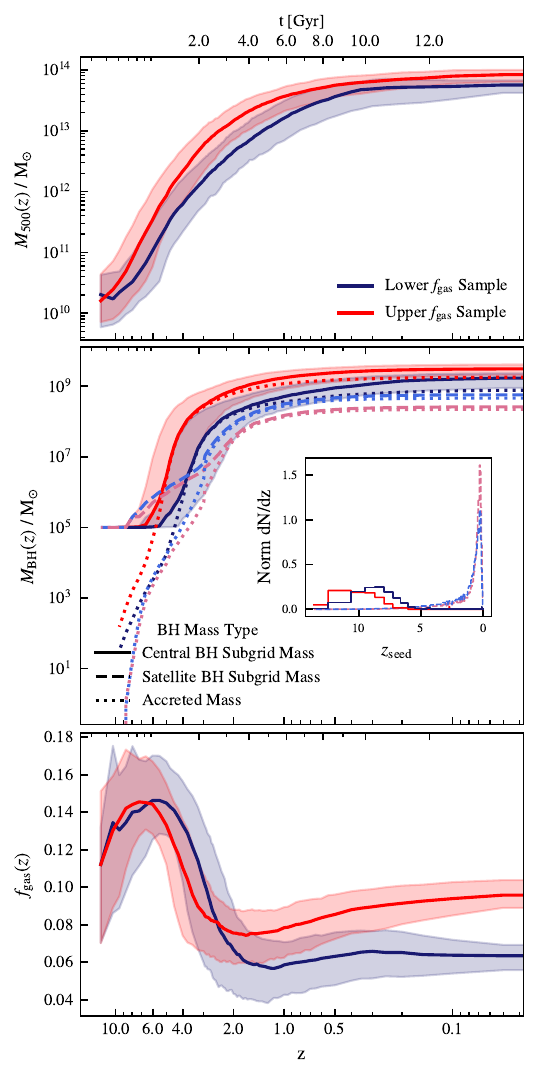}
 \caption{The redshift evolution of central BHs and their haloes within the $z=0$ mass range $\mathrm{13.75 \leq log_{10}(\mfive) \leq 14}$ split into the 90th (upper, red) and 10th (lower, navy) percentiles of the \fgas{} distribution at $z=0$. Solid lines show the median trend per sample, and shaded regions enclose the 16-84th percentile range. \textbf{(Top)} The evolution of median halo mass, \mfive{}, with redshift. \textbf{(Centre)} BH mass, \mbh{}, as a function of redshift. Solid lines show the median central BH masses for the upper (red) and lower (navy) \fgas{} samples, and dashed lines show the upper (pink) and lower (light blue) median satellite BH population masses. Dotted lines indicate the total accreted BH mass per sample. The inset panel shows the normalised number of BHs per redshift interval against the redshift at which these BHs were formed, $z_\text{seed}$. \textbf{(Bottom)} Gas fraction within \rfive{}, \fgas{}, as a function of redshift. Haloes that collapse earlier form central BHs earlier, leading to earlier gas expulsion and re-accretion, and resulting in a relatively high gas fraction compared to haloes of the same present-day mass that formed later.}
 \label{fig:BH_tracking_subplots}
\end{figure}

    Fig. \ref{fig:BH_tracking_subplots} displays the results of BH tracking for the upper and lower \fgas{} samples, shown in red and navy, respectively. Across all panels, the solid lines denote the median track of the 710 central BHs per sample or the halo properties that they're linked to, and the shaded regions enclose the 16-84th percentile range. 

    The top panel shows the redshift evolution of $M_{500}$ for the upper and lower samples. As defined, all haloes at $z = 0$ lie within the mass bin $10^{13.75} \msun \leq \mfive < 10^{14} \msun$. Within this selection, there is a small offset in median halo mass at $z = 0$ between samples, with the upper sample having a median $\approx 0.1$ dex higher. However, our conclusions are unaffected by this given the small magnitude of this difference and the overlap in the percentile ranges of the halo mass distributions of the two samples. Both sets of haloes begin their growth at $z \approx 10$, but the upper \fgas{} sample experiences a much quicker increase in mass, resulting in earlier BH growth. 

    The centre panel shows the evolution of BH mass across redshift for the upper and lower \fgas{} samples. Solid lines show the median central BH mass, and dashed lines show the median satellite BH population mass for the upper (pink) and lower (light blue) \fgas{} samples. The corresponding dotted lines show the total accreted masses of each BH subset, i.e., the contribution to the BH mass through accretion, rather than via mergers. We emphasise again that in this plot, we are tracking fixed BH populations of central and satellite BHs, rather than tracking fixed haloes. To visualise BH formation times, the inset panel shows the normalised number of BHs per redshift bin as a function of their formation redshift, $z_{\mathrm{seed}}$. 
    
    As shown in Fig.~\ref{fig:cent_sat_vertical_hex}, systems with high gas fractions at $z = 0$ have more massive central BHs. These upper \fgas{} systems formed their central BHs earlier, at $z \approx 11$, with their growth beginning at $z \approx 6$. As haloes in the lower \fgas{} sample grow comparatively more slowly at early times, the formation of their BHs is also delayed with respect to those in the higher \fgas{} sample.  These median formation times are $z \approx 9$ and $z \approx 5$ for the higher and lower \fgas{} samples, respectively. 

    The lower panel of Fig.~\ref{fig:BH_tracking_subplots} shows the evolution of halo gas fractions for the upper and lower \fgas{} samples. We see that the high \fgas{} systems eject approximately half of their gas from \rfive{} between $z \approx 6$ and $z \approx 3$, coinciding with the time of rapid central BH growth. Gas that was expelled from \rfive{} (but not far enough to escape the halo potential well altogether) is then re-accreted onto the halo from $z \approx 2$, when the gas fraction begins to increase. A similar (but delayed) trend can be seen for the lower \fgas{} sample, wherein feedback begins at $z \approx 5$ and re-accretion begins at $z \approx 1.5$. These systems experience significantly more quenching due to the ejection of gas at a fixed halo mass being more effective at later times (as halo binding energy $\phi \propto -M^{5/3}(1+z)$). Interestingly, the lower \fgas{} sample's gas fraction plateaus after $z \approx 0.4$ coinciding with a similar flattening in the halo mass growth. This suggests that the stunted re-accretion may be linked to the reduced deepening of the halo potential well at late times.
    
    For both samples, the formation of satellite BHs begins at $z \approx 5$, with the vast majority being formed below $z \approx 1$. While the formation epochs are similar, high \fgas{} haloes at $z = 0$ host about 50 per cent more satellite BHs, primarily because their higher halo masses naturally imply richer satellite populations. Despite this numerical advantage, satellite BHs in the lower \fgas{} sample begin growing in mass earlier and ultimately reach much higher combined masses, comparable to that of their central BHs. This anti-correlation between central and satellite BH mass may be explained by the growth histories of their host galaxies: satellite galaxies originate as centrals, form a BH, and evolve before being accreted into the group or cluster, where ram-pressure stripping truncates their growth. Satellites that fall in later can therefore grow their BHs for longer, potentially accounting for the higher satellite masses in the lower \fgas{} sample. Alternatively, these results may point to a direct physical connection between central and satellite BHs, suggesting some form of mutual regulation or quenching. We explore this possible “proximity effect” in Section~\ref{sec:proximity effect}.
    
\begin{figure}
 \includegraphics[width=\columnwidth]{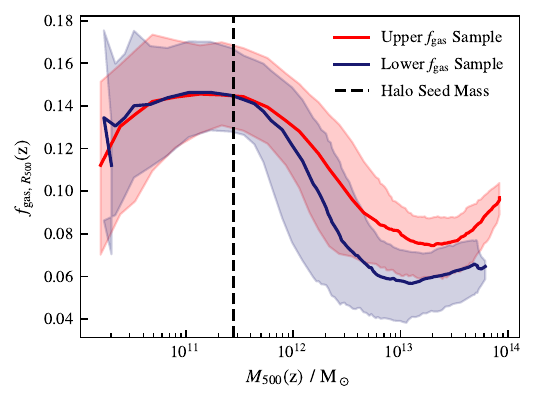}
 \caption{{The evolution of the halo gas fraction as a function of halo mass, \mfive{}, for haloes selected within the upper (red) and lower (navy) 10th percentiles of the\fgas{} distribution at $z=0$ within within ${13.75 \leq \text{log}_{10}(\mfive / \msun) \leq 14}$. For each BH/halo system, the main progenitor is tracked across redshift, producing an evolutionary track in the \fgas-\mfive plane. Solid lines show the median evolutionary track for each sample, computed in bins of \mfive{}, while the shaded regions enclose the 16th–84th percentile range of the track distributions. The vertical black dashed line marks the halo mass at which BHs are seeded in FLAMINGO, ${M}_\text{seed} = 2.757 \times 10^{11} \msun$. The depletion of gas occurs when haloes reach a characteristic mass threshold, but the redshift at which this occurs determines how efficiently that gas is expelled and subsequently re-accreted.}}
 \label{fig:BH_tracking_m500_fgas}
\end{figure}

    {The ejection and re-accretion of halo gas can be seen in Fig.~\ref{fig:BH_tracking_m500_fgas}, which shows the same upper and lower \fgas{} samples as before. The main progenitor haloes are tracked through the same redshift range as Fig.~\ref{fig:BH_tracking_subplots}. For each BH/halo system, the halo gas fraction and \mfive{} are recorded at each redshift, producing an evolutionary track in the \fgas–\mfive{} plane. The median evolutionary track for each sample is then computed from these individual tracks in bins of \mfive{} and plotted as a solid line. The shaded regions indicate the 16th–84th percentile range of the track distributions.} In both cases, gas ejection begins at halo masses above the scale where BHs are seeded (indicated by the vertical dashed line), but the redshift at which this occurs is not fixed. Instead, the halo formation history determines when the mass threshold for BH formation is crossed, and thus when feedback is triggered, which in turn sets both the timing and efficiency of gas ejection and re-accretion. \citet{booth_dark_2010, booth_towards_2011} and \citet{elbers_flamingo_2025} likewise found that the halo formation epoch regulates the onset of BH growth and the subsequent expulsion of gas. Exactly why the feedback is more/less effective if haloes cross this mass threshold at different times is an interesting question: later-triggered feedback may leave less time for re-accretion, while it may also be intrinsically more effective at low redshift when halo and cosmic gas densities are lower. Additionally, as the galaxy around the BH varies with redshift, the effects of stellar feedback may also be significant.

        \begin{figure}
     \includegraphics[width=\columnwidth]{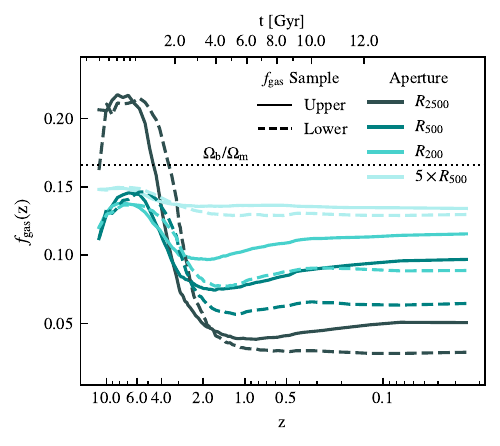}
     \caption{The gas fraction history as a function of redshift in haloes within the $z=0$ mass bin ${13.75 \leq \text{log}_{10}(\mfive / \msun)\leq 14}$ for different apertures: $f_{\mathrm{gas}, \ R_{2500}}(z)$, $f_{\mathrm{gas}, \ R_{500}}(z)$, $f_{\mathrm{gas}, \ R_{200}}(z)$, and $f_{\mathrm{gas}, \ 5 \times R_{500}}(z)$. Darker coloured lines represent the gas fraction within smaller apertures. The solid lines indicate the haloes within the 90th (upper) percentile of $f_{\mathrm{gas}}$ at $z = 0$ and the dashed lines the 10th (lower) percentile. Gas fractions decline sharply at the onset of AGN feedback, with smaller apertures showing earlier and stronger depletion. At late times, re-accretion begins from the outside-in, with larger apertures increasing in gas fraction earlier.} 
     \label{fig:fgas_apertures}
    \end{figure}

    To explore re-accretion further, we examine the redshift evolution of halo gas fractions within multiple apertures: $R_{2500}$, $R_{500}$, $R_{200}$, and $5 \times R_{500}$, with spherical overdensity quantities computed using SOAP. These results are shown in Fig.~\ref{fig:fgas_apertures}, with larger apertures indicated by progressively lighter shades for the upper (solid) and lower (dashed) \fgas{} samples.

    {Across all apertures, gas fractions initially decline at the onset of central BH feedback. Smaller apertures lose gas earlier and more severely, as these are closer to, and therefore more affected by, the central BH. For both samples, the strongest reduction occurs within $R_{2500}$, which loses approximately 80\% of its original gas content, and shifts from being the most gas rich aperture to the most gas poor. The fractional loss decreases with increasing aperture size, such that the ejection beyond $5 \times R_{500}$ is relatively modest.}
        
    {The lower-\fgas{} sample, whose central BH feedback begins later, consistently shows stronger depletion across all apertures. This provides further evidence that feedback is more effective at lower redshift, as discussed previously.}

    {   Signs of re-accretion are visible in all but the largest aperture. For both samples, the recovery begins first within $R_{200}$, followed by $R_{500}$ and then $R_{2500}$, consistent with an outside-in accretion scenario. Although the gas fraction within $5 \times R_{500}$ remains relatively unaffected, there is a large increase in the gas fraction within smaller radii, suggesting that some gas which has been ejected, but not beyond $5 \times R_{500}$, has fallen back in, providing indirect evidence for the re-accretion of previously ejected gas. Although re-accretion proceeds efficiently at intermediate redshifts, it appears to plateau at late times.  At this particular mass scale ($\mfive\sim10^{14}$ M$_\odot$), only a fraction of the original gas content is restored, showing that BH feedback causes long-lasting suppression. The largest aperture exhibits the weakest gas depletion and no obvious sign of re-accretion. Note that for the upper \fgas{} sample, most of this depletion is on account of star formation rather than gas ejection. }

    {
    To distinguish whether the observed recovery of gas is, in fact, a result of the re-accretion of previously ejected gas (as opposed to accretion of previously unassociated gas), we track the gas particles that reside within \rfive{} of haloes in the upper and lower \fgas{} samples at $z=0$. We broadly classify a gas particle as re-accreted if it previously resided within \rfive{} of a halo, was subsequently ejected beyond that radius, and later re-entered within \rfive{} of a halo.  Note that because of the hierarchical nature of halo assembly, asking whether a particle was ejected and re-accreted onto a system is a somewhat nuanced question.  Broadly speaking, one can define three re-accretion channels.  The first channel, which we label \textit{`Main $\rightarrow$ Main'} corresponds to gas which was originally within \rfive{} of a halo on the main progenitor branch, was ejected outside of \rfive{}, and was later re-accreted onto this same main progenitor halo.  We expect this to channel to be the dominant source of re-accreted gas.  The second channel, \textit{`Other $\rightarrow$ Main'}, describes gas that originally resided within \rfive{} of a halo outside of the main progenitor branch, was expelled, and subsequently accreted onto the main progenitor halo.  We still refer to this as re-accretion, since in the absence of feedback this gas would have ended up in the final system.  The final channel, \textit{`Other $\rightarrow$ Other'}, is associated with gas that originated within \rfive{} of a halo outside of the main progenitor branch, was expelled, and subsequently re-accreted onto another (or the same) halo outside of the main progenitor branch.  The gas in this channel is delivered to the final system via a merger onto the main branch.}
    
    {From our analysis we find the \textit{`Other $\rightarrow$ Other'} channel makes a negligible contribution to the final gas fraction of haloes.  The \textit{`Other $\rightarrow$ Main'} channel is more significant (contributing up to $\approx$ 10\%) of the final gas mass within \rfive{} but, based on how this channel is defined, it is difficult to distinguish gas that was once in another halo, ejected, and then accreted on the main progenitor branch from a scenario where gas is ram pressure stripped from an infalling satellite.  We therefore focus our analysis on the most significant (and unambiguous) channel, \textit{`Main $\rightarrow$ Main'}, where gas is ejected from and re-accreted onto the main progenitor branch.}  

    {Algorithmically, our approach is as follows for the \textit{`Main $\rightarrow$ Main'} channel.  We select all gas particles that lie within \rfive{} of systems in the upper and lower \fgas{} samples. We then trace these particles back in time using their unique particle IDs, extracting their properties from every available snapshot (at redshifts $z = 0, 0.1, 0.2, 0.3, 0.4, 0.5, 0.75, 1, 1.5, 2, 3, 4,$ and $5$). We use the \textsc{HBT-HERONS} merger trees for each halo in the \fgas{} samples and record whether or not a given gas particle is within \rfive{} of the main progenitor halo at each snapshot.  As already described, a particle is flagged as being re-accreted if it previously resided within \rfive{} of the main progenitor halo, was subsequently ejected beyond \rfive{}, and later re-entered within \rfive{} of the main progenitor.}

    \begin{figure}
     \includegraphics[width=\columnwidth]{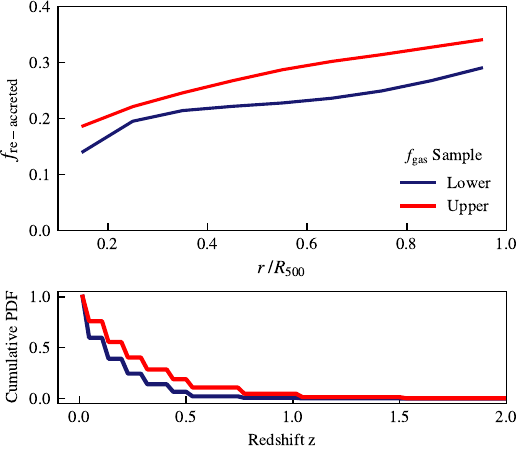}
     \caption{{\textbf{(Top)} The mass fraction of gas particles from the upper (red) and lower (navy) \fgas{} samples that have, at some epoch, been ejected from and subsequently re-accreted into \rfive{} of a halo on the main progenitor branch, $f_\text{re-accreted}$, as a function of radius, $r / \rfive{}$. The re-accretion fraction within main progenitor haloes increases with radius, consistent with an outside-in re-accretion scenario. \textbf{(Bottom)} The cumulative PDF (normalised to $z=0$) of the redshift of re-accretion; i.e., the redshift at which a gas particle is first identified as having re-entered \rfive{} of the main progenitor halo from which it was ejected at an earlier time. Gas particles in upper \fgas{} sample are re-accreted earlier than in the lower \fgas{} sample, consistent with Fig.\ref{fig:BH_tracking_subplots}.}}
     \label{fig:reaccretion_fraction_by_origin_and_sample}
    \end{figure}
    
    {The top panel of Fig.~\ref{fig:reaccretion_fraction_by_origin_and_sample} shows the mass fraction of gas particles that have been re-accreted into haloes in the main progenitor branch by $z=0$, $f_\text{re-accreted}$, in shells of \rfive{} for the upper (red) and lower (navy) \fgas{} samples. We note that, because of the limited number of snapshots and their wide cadence, these re-accretion fractions are likely an underestimate, as gas particles that are ejected from and subsequently re-accreted into a halo between consecutive snapshots are not captured by our analysis.}
    
    {%Main progenitor re-accretion dominates the total fraction at all radii and increases with radius, consistent with the outside-in re-accretion scenario shown in Fig.~\ref{fig:fgas_apertures}. 
    The maximum \freac{} reaches 34\% in the upper and 29\% in the lower \fgas{} sample. Across all radii, the upper \fgas{} sample exhibits a systematically higher \freac{} than the lower \fgas{} sample, reflecting the earlier onset of re-accretion in the former. The increase in the overall gas fraction due to main progenitor re-accretion is consistent with the increase in \fgas{} shown in the bottom panel of \ref{fig:BH_tracking_subplots}. %\textcolor{red}{We cannot distinguish \textit{Other $\rightarrow$ Main} re-accretion from ram pressure stripping, so we leave this for future work. Other $\rightarrow$ Other is negligible in comparison.} 
    This is also reflected in the bottom panel of Fig.~\ref{fig:reaccretion_fraction_by_origin_and_sample}, which shows the cumulative PDF (normalised to $z=0$) for the redshift of re-accretion
     i.e., the snapshot redshift at which a gas particle is first identified as having re-entered \rfive{} of the main progenitor halo from which it was ejected at earlier times. The earlier onset of re-accretion in the upper \fgas{} sample can be seen explicitly. The staircase pattern reflects the limited number of snapshots and their relatively wide temporal spacing. Nevertheless, the distribution remains informative for exploring the overall trend.}
      
    Returning to our earlier hypothesis, we find that haloes with above-average present-day gas fractions tend to have grown earlier in cosmic time. These early-forming haloes form their central BHs at higher redshifts, enabling prolonged accretion-driven growth. As a result, these BHs reach higher masses and become overmassive relative to the general population. By $z \approx 6$, strong AGN feedback from the central BH expels a significant fraction of the halo gas to large radii, and effectively quenches further accretion. While the central BH may continue to grow via mergers and remain overmassive, the growth of satellite BHs is suppressed, resulting in a larger population of low-mass satellites at $z = 0$. As the halo continues to evolve, previously ejected gas begins to re-accrete, leading to a rise in $\fgas$ around $z \approx 1.5$. As shown in Fig.~\ref{fig:fgas_apertures}, this re-accretion proceeds in an outside-in fashion. Evidence for re-accretion was also reported by \citet{lucie-smith_cosmological_2025}, who found that both the amount of gas recaptured and the redshift at which it occurs depend on halo mass and feedback strength in various FLAMINGO runs.
 
    The evolutionary pathway differs for haloes with below-average present-day gas fractions. These haloes assemble later than those in the upper sample, resulting in a delayed formation of their central BHs. Central BH growth begins around $z = 5$, coinciding with the appearance of satellite BHs. Due to the delayed onset, the central BHs reach lower masses and the satellites reach higher masses by the time significant feedback occurs. Nevertheless, this feedback is more impactful, as it occurs at a lower redshift, when gas densities and binding energies are smaller. Gas re-accretion is also delayed relative to the upper sample and is much less efficient, possibly due to accretion onto satellite galaxies in the halo outskirts.
    
    Understanding the role of satellite BHs in this context is non-trivial. In particular, it remains unclear how satellite BHs in the lower sample grow to significantly higher masses despite their lower numbers. We examine this in detail in Section~\ref{sec:proximity effect}, where we explore the radial distributions of these BH populations.

\section{The Proximity Effect}
\label{sec:proximity effect}

\begin{figure}
 \includegraphics[width=\columnwidth]{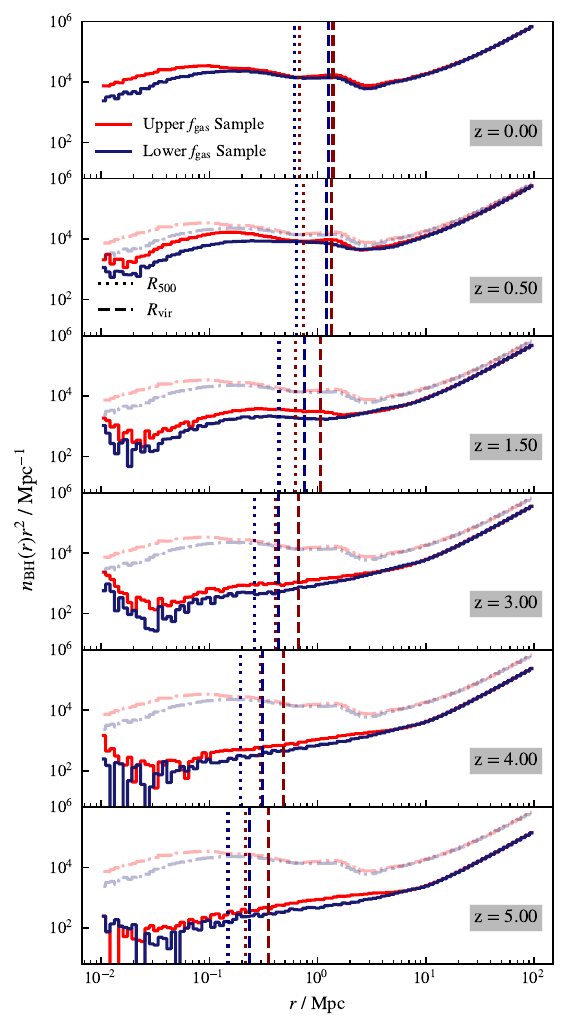}
 \caption{Radial number density profiles of BHs centred on the central BHs of the upper (red) and lower (navy) $z=0$ \fgas{} samples, shown at various redshifts. In the same colours, the median virial radius, \rvir, for each sample is indicated by the vertical dashed line, while \rfive{} is shown by the vertical dotted line. Faint dot-dashed lines in red and navy show the corresponding $z=0$ profiles for reference. Haloes with high gas fractions at $z=0$ have more satellite BHs across the entire redshift range.}
 \label{fig:n_density_profiles}
\end{figure}

\begin{figure}
 \includegraphics[width=\columnwidth]{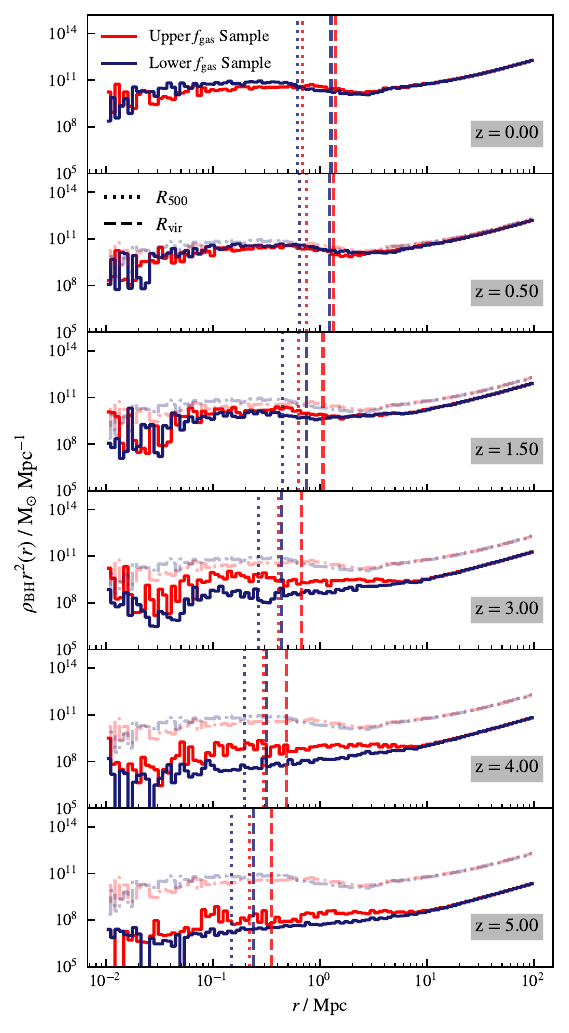}
 \caption{Radial mass density profiles of BHs centred on the central BHs of the upper (red) and lower (navy) $z=0$ \fgas{} samples, shown at various redshifts. The median virial radius, \rvir, for the upper and lower \fgas{} samples is indicated by the vertical dashed line, while \rfive{} is shown by the vertical dotted line, in red and navy, respectively. Faint dot-dashed lines in red and navy show the corresponding $z=0$ profiles for reference. Although the upper \fgas{} sample maintains a consistently higher BH \textit{number} density of satellites across all redshifts, it only dominates in BH mass density until $z=1.5$ (see Fig. \ref{fig:n_density_profiles}), after which the lower gas fraction sample overtakes. Despite hosting fewer satellites by number, the lower \fgas{} sample's BHs grow to be more massive by low redshift.}
 \label{fig:m_density_profiles}
\end{figure}

    In Fig. \ref{fig:cent_sat_vertical_hex}, we observed an anti-correlation between the scatter in central and satellite BH masses: haloes hosting overmassive central BHs tend to have undermassive satellite BH populations, and vice versa. Furthermore, Fig. \ref{fig:BH_tracking_subplots} showed that, within the mass bin ${10^{13.75} \msun \leq \mfive \leq 10^{14}\msun}$, haloes with high gas fractions and overmassive central BHs contain approximately 1.5 times as many satellite BHs as their low-\fgas{} counterparts. However, despite their greater number, the satellites in these high-\fgas{} haloes are significantly less massive compared to those in low-\fgas{} haloes. The latter grow to much higher masses, such that their total mass becomes comparable to that of the central BH by low redshift. This raises a key question: what suppresses satellite BH growth in the high-\fgas{} (upper) sample, or conversely, what drives the enhanced satellite BH growth in the low-\fgas{} (lower) sample?

    One possibility is that the relevant processes occur during preprocessing, before the satellite BHs (and their subhaloes) fall into groups. Although satellites in the upper and lower \fgas{} samples are formed at similar epochs (see Fig. \ref{fig:BH_tracking_subplots}), those in the lower \fgas{} sample may reside in subhaloes with later infall times, allowing them to accrete efficiently from a local gas supply for longer. By the time these subhaloes fall into the group environment and lose  gas through ram-pressure stripping, their BHs are already more massive. 
    
    Alternatively - or additionally, since the two effects need not be independent - these results may suggest a form of mutual quenching, wherein feedback from central BHs suppresses accretion onto surrounding BHs destined to become satellites of the main progenitor halo. In this scenario, the early, intense feedback from the upper \fgas{} sample central BHs inhibits the growth of their satellite BHs, producing haloes with many satellites but relatively little satellite BH mass. To explore these scenarios, we examine the redshift evolution of radial profiles centred on the central BH.

    For the same two central BH samples described before (comprising 710 BHs each from the upper and lower 10th percentiles of the \fgas{} distribution within the halo mass bin ${10^{13.75}\msun \leq \mfive \ \leq 10^{14} \msun}$ at $z=0$) we construct radial profiles centred on each central BH. Specifically, we define a sphere of radius 100 physical Mpc around each central BH and perform a nearest-neighbour search to identify all BHs within this region, excluding the central BH itself to avoid skewing average quantities. The resulting radial distributions from all spheres in a given sample are then aggregated. This procedure is repeated at each available snapshot across the redshift range $z = 0 - 5$.

    Fig.~\ref{fig:n_density_profiles} shows the radial number density profiles multiplied by $r^2$, $n(r)r^2$, for the upper (red) and lower (navy) \fgas{} samples, with each panel corresponding to a different redshift. In each panel, the $z=0$ profiles are shown as faint dot-dashed lines for reference. The dashed vertical line indicates the median virial radius of haloes in both samples, $R_{\mathrm{vir}}$, using the definition presented in \citealt{bryan_statistical_1998}, with the upper and lower \fgas{} samples shown in red and navy, respectively. In the same colours, the dotted line marks the median $R_{500}$ for each sample.     The upper \fgas{} sample consistently contains a higher number of BHs than the lower \fgas{} sample, even up to $z = 5$, with most forming after $z \approx 0.5$. As discussed earlier, this is likely due to these haloes being slightly more massive, and hence containing more satellites. In fact, the upper \fgas{} sample remains dominant in BH number density over the entire redshift range.
    
    Fig.~\ref{fig:m_density_profiles} shows the radial BH \textit{mass} density profiles multiplied by $r^{2}$, $\rho(r)r^{2}$, for both samples across the redshift range, following the same formatting as Fig.~\ref{fig:n_density_profiles}. Unlike the BH number density profiles, the BH mass is not always most concentrated near the central BH. At $z \approx 3$, we observe a suppression in the central mass density of the lower \fgas{} sample, which coincides with the onset of feedback in these haloes. This suppression begins to diminish by $z \approx 1.5$. Interestingly, although the upper \fgas{} sample has a larger BH number density across all redshifts, it only has a larger BH mass density until $z \approx 1.5$. After this point, the lower \fgas{} sample overtakes it in BH mass, indicating that although this sample has fewer satellite BHs by number, the combined mass of the satellite BHs becomes significant at low redshift.

    In addition to the number and mass density profiles, we have also examined the fraction of satellite BHs that are actively accreting gas (using their instantaneous accreting rates) as function of radius from the central BH and redshift.  Consistent with the mass density profile trends shown in Fig.~\ref{fig:m_density_profiles}, we find that a larger fraction of satellite BHs within a few Mpc of the central BH are actively accreting gas in the low \fgas{} sample compared to the upper \fgas{} sample below $z \approx 1.5$.
    
    While these trends have established the time frame over which the anti-correlation in gas fraction and satellite BH mass is established (i.e., below $z\approx1.5$) they do not allow us to definitively distinguish between the two physical scenarios (central BH quenching of satellites vs.~differences in the environmental processing of the satellites) discussed above.  We leave a more detailed analysis of these possibilities for future work.

\section{Summary and Conclusions}
\label{sec:summary}
The gas fractions of galaxy groups and clusters provide a key observational link between astrophysics and cosmology in large-scale structure. Several modern cosmological simulations and analytic models are calibrated against the median gas fraction–halo mass relation of massive systems, but this relation exhibits substantial intrinsic scatter that has not received much attention in the literature.  Understanding the physical drivers of this scatter is likely to be crucial for both interpreting observations and improving models of feedback. 

\citet{davies_gas_2019, davies_quenching_2020} previously identified a negative correlation between \fgas{} and central \mbh{} for haloes of $10^{11.5}\msun \leq M_{200} \leq 10^{13}\msun$. However, whether this result also extends to groups and clusters was unclear, due to the limited simulation volumes employed in those studies.  Furthermore, groups and clusters also host massive satellite BHs, whose combined masses - and therefore cumulative feedback - are comparable to that of their centrals.

The aim of this work was therefore not only to probe higher halo masses, but also to disentangle the relative roles of central and satellite BHs in shaping the gas content of massive haloes across cosmic time. To achieve this, we employed the high-resolution 1 Gpc-a-side FLAMINGO simulation, calibrated to reproduce observed median gas mass fraction--halo mass relation of low-z groups and clusters as well as the present-day galaxy stellar mass function. We examined the scatter around the \fgas-\mfive{} relation and its dependence on the total, central, and satellite BH population masses, before tracking a subset of BHs through cosmic time, to understand how their assembly histories have shaped their present-day gas content. Our main results are as follows:

\begin{itemize}
    \item Consistent with previous studies, we find a negative correlation between \fgas{} and central \mratio{} for haloes with $\mfive < 10^{12.5}\msun$ at $z=0$. At low halo masses, therefore, galaxies hosting overmassive central BHs reside in gas-poor haloes, while undermassive BHs reside in gas-rich haloes. This confirms that a major source of scatter in the \fgas--\mfive{} relation is due to AGN feedback from the central BH (Fig.~\ref{fig:cent_sat_vertical_hex}, top).  The scatter is driven by a form of assembly bias, whereby early-collapsing (and more concentrated) haloes form their central BHs faster leading to more efficient early gas ejection compared to later forming haloes of the same mass.
    
    \item For more massive haloes ($10^{13}\msun < \mfive < 10^{14.5} \msun$ at $z=0$), this trend reverses: \fgas{} correlates positively with central \mratio{}. Overmassive central BHs in this regime are associated with relatively gas-\textit{rich} haloes (Fig.~\ref{fig:cent_sat_vertical_hex}, top). 

    \item Satellite BH populations exhibit the opposite behaviour: low-mass, gas-poor haloes (and high-mass, gas-rich haloes) host undermassive satellite BH populations, and vice versa. This establishes a clear anti-correlation between central and satellite BH masses - haloes with an overmassive central BH are likely to host an undermassive population of satellite BHs (Fig.~\ref{fig:cent_sat_vertical_hex}, bottom). 

    \item Across almost all halo masses, central BHs dominate the dependence of \fgas-\mfive{} scatter on BH mass. Only in the most massive haloes ($\mfive \geq 10^{14.5} \msun$), where the combined mass of satellite BHs approaches that of the central BH (Fig.~\ref{fig:BH_populations}), does this correlation weaken, (Fig.~\ref{fig:tot_hex}). Whether this reflects weaker feedback in deep potential wells, significant satellite BH feedback, or limited statistics remains unclear.  

    \item The correlation between \fgas{} and central \mratio{} evolves with redshift. At $z > 1.5$ (maximum $\mfive \approx 10^{13.6} \msun$), the correlation is uniformly negative: overmassive (undermassive) BHs reside in gas-poor (gas-rich) haloes regardless of halo mass (Fig.~\ref{fig:multiz_horizontal}, top left). At $1.5 \leq z < 0.25$ (maximum $\mfive \approx 10^{14.8} \msun$), the trend reverses for $\mfive > 10^{13} \msun$, with overmassive BHs in group-scale haloes increasingly linked to high gas fractions. By $z=0.75$, the relation is degenerate, with overmassive BHs found in both extremely gas-rich and gas-poor haloes (Fig.~\ref{fig:multiz_horizontal}, top right). By $z=0$, the reversal is complete: above (below) $\mfive = 10^{13}\msun$, overmassive (undermassive) BHs are hosted by gas-rich (gas-poor) haloes (Fig.~\ref{fig:multiz_horizontal}, bottom right).

    \item By tracking progenitor haloes (within $10^{13.75}\msun \leq \mfive (z=0) < 10^{14}\msun$) and BHs with extreme gas fractions at $z=0$ through cosmic time, we probe this surprising reversal and find that haloes with above-average present-day gas fractions assembled earlier in time (Fig.~\ref{fig:BH_tracking_subplots}, top), so form their central BH at higher redshift, enabling prolonged accretion-driven growth such that they become overmassive relative to the general BH population (Fig.~\ref{fig:BH_tracking_subplots}, centre). An earlier onset of strong AGN feedback then expels around half of the halo gas from \rfive{} (Fig.~\ref{fig:BH_tracking_subplots}, bottom) as well as beyond $5 \times \rfive{}$ (Fig.~\ref{fig:fgas_apertures}). At $z=1.5$, outside-in re-accretion begins (Fig.~\ref{fig:reaccretion_fraction_by_origin_and_sample}, Fig.~\ref{fig:fgas_apertures}, Fig.~\ref{fig:BH_tracking_subplots}, bottom), where previously ejected gas begins to re-accrete, driving their recovery to high gas fractions. 
    
    \item On the other hand, haloes with below-average present-day gas fractions assemble later (Fig.~\ref{fig:BH_tracking_subplots}, upper), delaying the formation and growth of their central BHs (Fig.~\ref{fig:BH_tracking_subplots}, centre) and allowing more time for satellite BHs to grow. Central BH feedback at lower redshift depletes a higher amount of gas (Fig.~\ref{fig:BH_tracking_subplots}, bottom), and re-accretion is not as efficient, resulting in a below-average gas fraction at the present-day (Fig.~\ref{fig:fgas_apertures}). 

    \item Our results show that the resulting halo gas fractions depend on the early formation history of the halo and its BHs. In particular, haloes that form earlier also have earlier forming BHs that expel gas from the main progenitor system earlier.  However, such early forming systems ultimately (by $z=0$) tend to be less efficient at expelling gas, possibly because there is more time for re-accretion compared to later forming haloes which expelled their gas more recently (Fig.~\ref{fig:BH_tracking_m500_fgas}). 

\end{itemize}

In the final stages of preparing this article, \citet{marini_impact_2025} presented an analysis of the X-ray detectability of galaxy groups in the relatively large Magneticum \textit{Box2/hr} run, which is 500 Mpc on a side. In particular, they explored the link between BH feedback, hot gas fractions, and X-ray detectability.  Interestingly, at $z=0$ they found an anti-correlation between BH mass and hot gas fraction over the entire mass range ($10^{12.5} \msun \leq \mfive < 10^{14.5} \msun$) they explored; see their Fig. 9.  On the other hand, our study using the FLAMINGO simulations shows a reversal of this trend above $\mfive{} \approx10^{13.0} \msun$.  The origin of this difference in findings is unclear, but may stem from differences in the AGN feedback prescriptions of the two simulations.  In this regard, it is interesting to note that Magneticum has a dual model for AGN feedback, where the efficiency of feedback is strongly elevated at low accretion rates (which occurs predominantly in more massive systems and at later times). In FLAMINGO, the feedback efficiency is independent of accretion rate. It is particularly interesting that their systems with relatively lower gas fractions at $z=0$ are continuing to eject substantial quantities of gas even at the present day (see their Fig. 5) while haloes with relatively higher gas fractions are in a re-accretion phase.  This differs from FLAMINGO, where both the lower and upper \fgas{} samples we have defined are firmly in a re-accretion phase and have been so since at least $z\approx1.5$. It would be instructive to perform a more direct comparison between these and other simulations, which we leave for future work.

In summary, we find that the scatter in the present-day \fgas–\mfive{} relation at the group and cluster scale is strongly influenced by the early growth history of central BHs and their host haloes. While the importance of high-redshift gas ejection in shaping present-day gas fractions is not unexpected, our results highlight how the formation history and growth of BHs play a decisive role in determining large-scale baryonic structure. 
Testing these ideas observationally will require joint measurements of BH masses and halo gas fractions, ideally in statistically large samples. Upcoming and ongoing efforts from surveys such as MASSIVE, which directly measures central BHs in the most massive nearby galaxies, to future high-resolution X-ray and Sunyaev-Zel'dovich programmes with facilities such as eROSITA, will provide crucial data for confronting these predictions and disentangling the roles of central and satellite BHs in shaping halo gas content.

\section*{Acknowledgements}
{The authors thank the anonymous referee for their helpful comments that improved the paper.} The authors would like to thank Jonah Conley, Rob Crain, Jonathan Davies, Megan Donahue, Ilaria Marini, Marie Martig, and Mark Voit for helpful discussions. EEC acknowledges an STFC PhD studentship at the LIV.INNO Centre for Doctoral Training “Innovation in Data Intensive Science”. This work was supported by the Science and Technology Facilities Council (grant number ST/Y002733/1). This project has received funding from the European Research Council (ERC) under the European Union’s Horizon 2020 research and innovation programme (grant agreement No 769130).
This work used the DiRAC@Durham facility managed by the Institute for Computational Cosmology on behalf of the STFC DiRAC HPC Facility (www.dirac.ac.uk). The equipment was funded by BEIS capital funding via STFC capital grants ST/K00042X/1, ST/P002293/1, ST/R002371/1 and ST/S002502/1, Durham University and STFC operations grant ST/R000832/1. DiRAC is part of the National e-Infrastructure.

%%%%%%%%%%%%%%%%%%%%%%%%%%%%%%%%%%%%%%%%%%%%%%%%%%
\section*{Data Availability}
A detailed description of the relevant FLAMINGO data products
and access arrangements is given in \citealt{helly_flamingo_2026}. The data supporting the plots within this article are available on reasonable request to the corresponding author.


\bibliographystyle{mnras}
\bibliography{reference} % if your bibtex file is called example.bib

%%%%%%%%%%%%%%%%%%%%%%%%%%%%%%%%%%%%%%%%%%%%%%%%%%

%%%%%%%%%%%%%%%%% APPENDICES %%%%%%%%%%%%%%%%%%%%%

%%%%%%%%%%%%%%%%%%%%%%%%%%%%%%%%%%%%%%%%%%%%%%%%%%

% Don't change these lines
\bsp	% typesetting comment
\label{lastpage}
\end{document}